\documentclass[12pt]{article}
\usepackage{latexsym}
\usepackage{graphicx}
\usepackage{times}

\begin{document}
\title{Superseded version of the WKB approximation and explanation of emergence of classicality}
\author{Wang Guowen\\College of Physics, Peking University, Beijing, China}
\date{December 25, 2006}
\maketitle

\begin{abstract}
Regarding the limit $\hbar \rightarrow 0$ as the classical limit
of quantum mechanics seems to be silly because $\hbar$ is a
definite constant of physics, but it was successfully used in the
derivation of the WKB approximation. A superseded version of the
WKB approximation is proposed in the classical limit $\alpha
\rightarrow 0$ where $\alpha=m/M$ is the screening parameter of an
object in which \emph{m} is the mass of the effective screening
layer and \emph{M} the total mass. This version is applicable to
not only approximate solution of Schr\"{o}dinger equation of a
quantum particle but also that of a nanoparticle. Moreover, the
version shows that the quantization rules for nanoparticles can be
achieved by substituting $\alpha\hbar$ for $\hbar$ in the
Bohr-Sommerfeld quantization rules of the old quantum theory. Most
importantly, the version helps clarify the essential difference
between classical and quantum realities and understand the
transition from quantum to classical mechanics as well as quantum
mechanics itself.
\end{abstract}

\section{Introduction}
The WKB (Wentzel-Kramers-Brillouin) approximation or
phase-integral approximation plays an important role in the
solution of Schr\"{o}dinger equation in the case where a particle
has low momentum and moves through a slowly varying potential and
in the proof of the Bohr-Sommerfeld quantization rules.[1-3] This
approximation method is based on the limit $\hbar\rightarrow 0$
which was first considered as the classical limit of quantum
theory by Max Planck who stated:$``$The classical theory can
simply be characterized by the fact that the quantum of action
becomes infinitesimally small.$"$[4] However, the fact that the
diffraction and interference of, for example, the grains of sand
do not really occur when they pass through slits can be properly
explained by considering that the outer matter of a tiny grain of
sand screens nearly completely the associated wave by the inner
matter.[5] This proposed screening effect gives a logical
description of transition from a quantum particle to a classical
object and gives a general momentum-position uncertainty relation
[6]:
\begin{equation}
\label{eq1} \triangle p_x^{(\alpha)}\triangle x=\alpha \triangle
p_x\triangle x\geq\frac{\alpha\hbar}{2}, \mbox{ }1\geq\alpha>0
\end{equation}
where $\alpha$ is the screening mass parameter defined by
$\alpha=m/M$ in which \emph{m} is the mass of the effective
screening layer and \emph{M} the total mass. For a spherical
nanoparticle shown in Fig.1, the corresponding screening size
parameter is $\sigma=r_m/r_M=1-(1-\alpha)^{1/3}$ [6]. The
thickness of the effective screening layer having quantum behavior
is estimated to be a few nanometers.
\begin{figure}[htbp]
\centerline{\includegraphics[width=1.6in,height=1.6in]{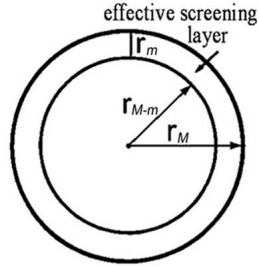}}
\label{Fig.1} \caption{Sketch of the effective screening layer in
a spherical nanoparticle.}
\end{figure}
Obviously the Heisenberg uncertainty relation is only applicable
to quantum particles ($\alpha$=1), but not applicable to
macroscopic objects ($\alpha\rightarrow 0$) and mesoscopic objects
(between). The fact that the momentum and position of a
macroscopic object are measurable simultaneously with finite
errors implies the limit $\alpha \rightarrow0$ instead of the
formal limit $\hbar \rightarrow 0$. It seems to be silly to regard
the universal fundamental physical constant $\hbar$ as a variable
quantity. We will thus propose a superseded version of the WKB
approximation in the classical limit $\alpha \rightarrow0$. This
limit is proper and general, which, as will be seen below,
corresponds to the emergence of classicality from quantum
mechanics.

\section{Superseded version of the WKB approximation}
Traditionally, the wave function in the WKB approximation is
expanded in terms of $\hbar$, which is based on the Planck limit
$\hbar\rightarrow 0$. Indeed, $\hbar=1.05457\times10^{-34}$
Joule-sec and the Planck limit has no physical interpretation.
Now, using the limit $\alpha \rightarrow0$ instead of the Planck
limit for derivation of the approximation, we assume that the
Schr\"{o}dinger equation of an object with an effective screening
parameter $\alpha$ moving through a one-dimensional potential
$V(x)$ is expressed by
\begin{equation}
\label{eq2}\frac{\mbox{d}^2\psi}{\mbox{d
}x^2}+\frac{2M}{\alpha^2\hbar^2}(E-V)\psi=0
\end{equation}
where $1\geq\alpha>0$. This equation implies that the momentum
operator now is $-i\alpha\hbar \partial/\partial x$
($-i\alpha\hbar \nabla$ in 3 dimensions). Of course the energy
operator correspondingly becomes $i\alpha\hbar \partial/\partial
t$. The equation also implies that the commutator
$[\hat{x},\hat{p}]=i\alpha \hbar \{x,p\}=i\alpha \hbar$ in which
$\{x,p\}$ is the Poisson bracket of \emph{x} and \emph{p} and that
the Bohm quantum potential [7] approaches 0 when
$\alpha\rightarrow 0$ instead of $\hbar\rightarrow 0$. The
approximate solution of the equation can be written in the form:
\begin{equation}
\label{eq3} \psi=\exp(\frac{i}{\alpha\hbar}\int^xy\mbox{d}x),
\mbox{ }y=y(x,\alpha)
\end{equation}
which satisfies
\begin{equation}
\label{eq4} (-\frac{y^2}{\alpha^2\hbar^2})\psi
+(\frac{i}{\alpha\hbar})y'\psi
+\frac{2M}{\alpha^2\hbar^2}(E-V)\psi=0
\end{equation}
where the prime denotes differentiation with respect to \emph{x}.
So, assuming $\psi\neq 0$, we obtain
\begin{equation}
\label{eq5} (\frac{\alpha\hbar}{i})y'=p^2-y^2, \mbox{ }p^2=2M(E-V)
\end{equation}
We now expand $y$ as a series in powers of $\alpha/i$:
\begin{equation}
\label{eq6}
y=y_0+(\frac{\alpha}{i})y_1+(\frac{\alpha}{i})^2y_2+\cdots, \mbox{
} 1\geq\alpha>0
\end{equation}
Therefore we have
\begin{equation}
\label{eq7}
(\frac{\alpha\hbar}{i})y'=(\frac{\alpha\hbar}{i})\sum^{\infty}_{n=0}(\frac{\alpha}{i})^ny'_n
\end{equation}

\begin{equation}
\label{eq8}
p^2-y^2=p^2-\sum^{\infty}_{l,m=0}(\frac{\alpha}{i})^{l+m}y_ly_m
\end{equation}
Thus Eq.5 becomes
\begin{equation}
\label{eq9} \hbar\sum^{\infty}_{n=1}(\frac{\alpha}{i})^ny'_{n-1}
=p^2-\sum^{\infty}_{l,m=0}(\frac{\alpha}{i})^{l+m}y_ly_m
\end{equation}
Equating coefficients of the same powers of $\alpha/i$ on the left
and right sides of the above equation, we get
\begin{equation}
\label{eq10} y_0=\pm p=\pm\sqrt{2M(E-V)}
\end{equation}

\begin{equation}
\label{eq11} \hbar y'_{n-1} =-\sum^{n}_{m=0}y_{n-m}y_m, \mbox{ }
n=1,2,3,\cdots
\end{equation}
From Eq.10 and Eq.11 we have
\begin{equation}
\label{eq12} y'_0=\pm\frac{1}{2}\cdot\frac{-2M
V'}{\sqrt{2M(E-V)}}=-\frac{MV'}{y_0}
\end{equation}

\begin{equation}
\label{eq13} \hbar y'_0=-2y_1y_0
\end{equation}
and hence get
\begin{equation}
\label{eq14} y_1=\frac{\hbar V'}{4(E-V)}
\end{equation}
Furthermore, we have the equations
\begin{equation}
\label{eq15} \hbar y'_1=-y^2_1-2y_0y_2
\end{equation}

\begin{equation}
\label{eq16}
y'_1=\frac{\hbar}{4}\cdot\frac{V''(E-V)+V'^2}{(E-V)^2}
\end{equation}
and hence obtain
\begin{equation}
\label{eq17} y_2=\frac{1}{2y_0}(-y^2_1-\hbar
y'_1)=\mp\frac{\hbar^2}{32}\cdot
\frac{5V'^2+4V''(E-V)}{(2M)^{1/2}(E-V)^{5/2}}
\end{equation}
Similarly from
\begin{equation}
\label{eq18} \hbar y'_2=-2y_3y_0-2y_2y_1
\end{equation}
it follows that
\begin{equation}
\label{eq19} y_3=\frac{1}{2y_0}(-\hbar
y'_2-2y_2y_1)=-\hbar\frac{\mbox{d}}{\mbox{d}x}(\frac{y_2}{2y_0})
\end{equation}
And similarly for higher terms in $\alpha/i$. Now we write the
power series as follows
\begin{eqnarray}
\label{eq20}y(x,\alpha)
=\pm\sqrt{2M(E-V)}+(\frac{\alpha}{i})\frac{\hbar V'}{4(E-V)} \nonumber \\
\mp(\frac{\alpha}{i})^2(\frac{\hbar^2}{32}\cdot
\frac{5V'^2+4V''(E-V)}{(2M)^{1/2}(E-V)^{5/2}})+\cdots
\end{eqnarray}
This formulas is the same as that obtained from the original
version of the WKB approximation when $\alpha=1$. Evidently, this
superseded version is general in the sense that it is applicable
to not only quantum particles ($\alpha=1$) but also nanoparticles
($1>\alpha>0$).

As well known, usually it is only necessary to take the first two
terms of the above series. By using the equation
\begin{equation}
\label{eq21} \int^x \frac{V'}{4(E-V)} \mbox{d}x=-\frac{1}{2}\log p
\end{equation}
we can thus write
\begin{equation}
\label{eq22} \psi=\frac{C_1}{\sqrt{p}}
\exp(\frac{i}{\alpha\hbar}\int^xp\mbox{d}x)+\frac{C_2}{\sqrt{p}}
\exp(-\frac{i}{\alpha\hbar}\int^xp\mbox{d}x), \mbox{ }
p=\sqrt{2M(E-V)}
\end{equation}
in which the coefficients $C_1$ and $C_2$ are determined by
boundary and normalization conditions. The difficulty that rises
is that the WKB approximation becomes inapplicable at the
classical returning points where $E-V$=0, but the approximate wave
functions near the points can be easily obtained from solving the
Schr\"{o}dinger equation and hence the connection formula between
the WKB wave functions and the approximate wave functions can be
derived in the way shown in many textbooks on quantum mechanics,
such as that by Merzbacher [8]. This version shows that the
quantization rules for nanoparticles can be achieved by
substituting $\alpha\hbar$ for $\hbar$ in the Bohr-Sommerfeld
quantization rules of the old quantum theory.

In order to investigate the relation between classical and quantum
mechanics, we now write Eq.3 as the following form:
\begin{equation}\label{eq23}
\psi=\exp(\frac{iS_\alpha}{\alpha\hbar}), \mbox{ }
S_\alpha=\int^xy(x,\alpha)\mbox{d}x
\end{equation}
Substituting this wave function into Eq.2 and assuming $\psi\neq
0$, we get
\begin{equation}\label{eq24}
(\frac{\mbox{d }S_\alpha }{\mbox{d
}x})^2+(\frac{\alpha\hbar}{i})\frac{\mbox{d}^2S_\alpha}{\mbox{d
}x^2}-2M(E-V)=0
\end{equation}
When $\alpha\rightarrow 0$, it becomes the well known
Hamilton-Jacobi equation
\begin{equation}\label{eq25}
(\frac{\mbox{d }S_0}{\mbox{d }x})^2-2M(E-V)=0
\end{equation}
in which $S_0$ is the action function of classical mechanics, so
the limit $\alpha\rightarrow 0$ corresponds to the emergence of
classicality from quantum mechanics. This point is vital for
understanding quantum mechanics.

\section{Conclusion}
A superseded version of the WKB approximation has been proposed in
the classical limit $\alpha \rightarrow 0$ where $\alpha=m/M$ is
the screening parameter of an object in which \emph{m} is the mass
of the effective screening layer and \emph{M} the total mass. This
version is applicable to not only approximate solution of
Schr\"{o}dinger equation of a quantum particle but also that of a
nanoparticle. Moreover, the version shows that the quantization
rules for nanoparticles can be achieved by substituting
$\alpha\hbar$ for $\hbar$ in the Bohr-Sommerfeld quantization
rules of the old quantum theory. Most importantly, the version
helps clarify the essential difference between classical and
quantum realities and understand the transition from quantum to
classical mechanics as well as quantum mechanics itself.

\end{document}